\documentclass[authoryear,preprint,12pt]{elsarticle}

\usepackage{graphics}

\usepackage{amssymb}

\usepackage[nodots]{numcompress}

\usepackage{lineno}

\usepackage{longtable}

\journal{Icarus, accepted}

\begin{document}

\begin{frontmatter}


\title{Atlas of three body mean motion resonances in the Solar System}

\author{Tabar\'e Gallardo\corref{cor1}}
\ead{gallardo@fisica.edu.uy}

\cortext[cor1]{Corresponding author}

\address{Departamento de Astronom\'{i}a, Instituto de F\'{i}sica, Facultad
de Ciencias, Universidad de la Rep\'{u}blica, Igu\'{a} 4225, 11400 Montevideo, Uruguay}

\begin{abstract}

We present a numerical method to estimate the strengths of arbitrary three body
mean motion resonances between two planets in circular coplanar orbits and a massless particle in an arbitrary
orbit. This method allows us to obtain an atlas of the three body
resonances in the Solar System showing  where are located and how strong are thousands of resonances
involving all the planets from 0 to 1000 au. This atlas confirms the dynamical relevance of the three body resonances involving Jupiter and Saturn in the asteroid belt but also shows the existence of a family of relatively strong three body resonances involving Uranus and Neptune
in the far Trans-Neptunian region and relatively strong resonances involving terrestrial and jovian planets in the inner planetary system.
We calculate the density of relevant resonances along the Solar System resulting that
the main asteroid belt is located in a region of the planetary system with the
lowest density of three body resonances.
The method  also allows the location of the equilibrium points showing the existence of asymmetric librations ($\sigma \neq 0^{\circ}$ or $180^{\circ}$).
We obtain the functional dependence of the resonance's strength with the order of the resonance and the eccentricity and inclination of the particle's orbit.
We identify some objects evolving in or very close to three body resonances with Earth-Jupiter, Saturn-Neptune and Uranus-Neptune apart from Jupiter-Saturn,
in particular the NEA 2009 SJ18 is evolving in the resonance 1-1E-1J and
the centaur 10199 Chariklo is evolving under the influence of the resonance 5-2S-2N.

\end{abstract}


\end{frontmatter}



\bigskip

%
%
%
%
%

\section{Introduction}
\label{intro}

Fifteen years ago,
concentrated over a period of about a year, a succession of papers were published \citep{mhp98,nm98,nm99,mn99} showing
 an intense theoretical and numerical work on three body mean motion resonances (TBRs)  involving an asteroid and two massive planets.
These papers, that have their roots on earlier works devoted to the study of an asteroid in zero order TBRs
\citep{wi33,ok35,ak88},
stated the relevance that the TBRs involving Jupiter-Saturn and also Mars-Jupiter have in the long term stability
 in the asteroid's region. In spite of being weaker than the two body resonances they are much more numerous
 generating several dynamical features in the asteroidal population, like concentrations for some values of semimajor axes, anomalous amplitude librations and chaotic evolutions \citep{nm98}. In particular, both borders of the main asteroid belt exhibit chaotic diffusion due to the superposition
 of several weak two-body and TBRs \citep{mh97,mn99}.
More recently, \citet{ss13} in a massive numerical
integration of 249,567 asteroids by $10^{5}$ years and looking at the time evolution of the critical angles,
identified thousands  of asteroids in TBRs with Jupiter and Saturn, concluding there
are more asteroids in TBRs than in two body resonances.

The approximate nominal position in semimajor axis of the TBRs taking arbitrary pairs of planets is very simple if we
ignore the secular motion of the perihelia and nodes of the three bodies. When these slow secular motions are taken into account
each of the nominal TBRs split in a multiplet of resonances all them very near the nominal one \citep{mo02}.
The challenge is to obtain the strength, width or
libration timescale
that give us the dynamical relevance of these resonances.
Analytical planar theories developed by  \citet{mhp98} and  \citet{nm99} allowed to describe and understand the dynamics
of the TBRs involving Jupiter and Saturn in the asteroidal region.
These theories are appropriated to study in detail specific
resonances with Jupiter and Saturn but its application
to any arbitrary resonance involving
any planet is not trivial,
which possibly explains the absence of papers published on this topic
since these years with the exceptions of a few ones devoted to specific scenarios \citep{gu05,de08,ca10,qu11}.

In this paper, from a different approach, we will obtain a global view of all dynamically relevant TBRs involving all the planets taken
by pairs along all the Solar System. Our method is not analytical but numerical and it is based on an estimation of
the strength of the resonances
which is obtained evaluating
 the effects of the mutual perturbations in all possible spatial configurations  of the three bodies satisfying the resonant condition.
It allows us to appreciate the effect of arbitrary eccentricity and inclination of the massless particle's orbit on the
resonance's strength but, in order to reduce the number of parameters involved in the problem,
we impose coplanar and circular orbits for the two perturbing planets. Nevertheless, the method can be extended to arbitrary planetary orbits.
 In the next section we start describing our numerical method  and we explore
 how the calculated strengths depend on the parameters of the problem.
In section \ref{atlas} we analyze the distribution of TBRs along the Solar System and discuss its dynamical effects
providing some examples. In section \ref{conc} we present the conclusions.

\section{Numerical approximation to the disturbing function for three body resonances}
\label{distur}

Three body resonances between a massless particle with an arbitrary orbit given by $(a_0,e,i,\Omega_0,\varpi_0)$ and two planets $P_1$ and $P_2$ in circular coplanar orbits occur when the critical angle
\begin{equation}\label{sigma}
    \sigma = k_0\lambda_0 +k_1\lambda_1 +k_2\lambda_2 - (k_0+k_1+k_2)\varpi_0
\end{equation}
is oscillating over time, being $\lambda_i$ the mean longitudes and $k_i$ integers. More precisely,
for massless particles with inclined orbits there are other possible definitions for
the critical angle involving combinations of its $\Omega_0$ and $\varpi_0$, but
it is possible to show that after an appropriate averaging procedure
the leading term in the expansion of the resulting disturbing function will be the one whose argument is given by Eq. (\ref{sigma}) \citep{nm99}.
In this work we will use the following notation: the order of the resonance is $q=|k_0+k_1+k_2|$, we call $p=|k_0| + |k_1| +|k_2| $
and we note $k_0+k_1P_1+k_2P_2$ the resonance involving planets $P_1$ and $P_2$.
The approximate nominal location of the resonance assuming unperturbed Keplerian motions is
\begin{equation}\label{nominal}
    a_0^{-3/2} \simeq -\frac{k_1}{k_0}\sqrt{(1+m_1)/a_1^3} -\frac{k_2}{k_0}\sqrt{(1+m_2)/a_2^3}
\end{equation}
where $a_i$ and $m_i$ are the mean semimajor axes of the planets and its masses expressed in solar masses respectively. In the real Solar System the actual location depends on the precession of the perihelia and the gravitational effects of the other planets not taken into account. In the very far Trans-Neptunian Region (TNR), depending on the resonance, the actual location could be shifted something between 0.1 au and 1 au.

According to \citet{nm99}, in the planar problem the Hamiltonian for the particle can be expressed in a simplified form as depending on canonical variables that depend on $(a,\sigma)$. The width in au of the resonance is $\Delta a \propto a^{3/2}\sqrt{\beta}$
where $\beta$ is the semi-amplitude of the resonant disturbing function, that we note as  $\mathfrak{R}(\sigma)$. After a thorough analytical procedure they
obtained an analytical expression for $\mathfrak{R}(\sigma)$ for the planar eccentric case asteroid-Jupiter-Saturn allowing them to obtain analytical solutions.
In this paper we are looking for a numerical approximation to $\mathfrak{R}(\sigma)$ for massless bodies in arbitrary
orbits in resonance with arbitrary pairs of planets considered in coplanar circular orbits. This approximation
will help us to understand which they are and how are distributed the dynamically relevant TBRs in the Solar System. The algorithm devised
here can be extended to eccentric planetary orbits and applied to other planetary systems.

The mean resonant disturbing function, $\mathfrak{R}(\sigma)$, that drives the resonant motion of the particle could be ideally calculated eliminating the short period terms of the
resonant disturbing function $R$ by means of
\begin{equation}\label{tiempo}
  \mathfrak{R}(\sigma) =  \frac{1}{T}\int_{0}^{T}
    R(\vec{r_0}(t),\vec{r_1}(t),\vec{r_2}(t)) dt
\end{equation}
where $T$ is an ideal interval, that means
long enough
 for the system to be evaluated in all possible configurations of the heliocentric positions, $\vec{r_i}$, of the three bodies but not long enough
to appreciate changes in $\sigma$. If we can admit the use of the Keplerian unperturbed positions  then we can substitute the integral in time domain by the
integral in phase space:
\begin{equation}\label{doble}
  \mathfrak{R}(\sigma) =  \frac{1}{4\pi^2}\int_{0}^{2\pi}d\lambda_1\int_{0}^{2\pi}
    R\Bigl(\lambda_0(\sigma,\lambda_1,\lambda_2,\varpi_0),\lambda_1,\lambda_2\Bigr)  d\lambda_2
\end{equation}
where $\lambda_0$ was explicitly written in terms of the variables $\lambda_1,\lambda_2$ and the parameters $\sigma,\varpi_0$ using Eq. (\ref{sigma})
and
where $R(\lambda_0,\lambda_1,\lambda_2) = R_{01} + R_{02}$
being
\begin{equation}
\label{Rij}
    R_{ij}= k^{2} m_{j} ( \frac{1}{r_{ij}} - \frac{\vec{r_i} \cdot \vec{r_j}}{r_j^3} )
\end{equation}
where $k$ is the Gaussian constant and $\vec{r_i}, \vec{r_j}$ are the heliocentric  positions of bodies with subindex $i$ and $j$ respectively.
Note that for each set of values $(\sigma,\lambda_1,\lambda_2,\varpi_0)$ there are $k_0$ values of $\lambda_0$ that satisfy Eq. (\ref{sigma}),
which are:
\begin{equation}\label{lambdas2}
    \lambda_0 =  \left(\sigma - k_1\lambda_1 -k_2\lambda_2 + (k_0+k_1+k_2)\varpi_0\right)/k_0 + n 2 \pi / k_0
\end{equation}
with $n=0,1,...,k_0-1$. All them contribute to $\mathfrak{R}(\sigma)$ in Eq. (\ref{doble}) so we have to evaluate all these $k_0$ terms and calculate the mean,
which is equivalent to integrate in $\lambda_0$ maintaining the condition (\ref{sigma}).
Up to this point, this scheme is analogue to the one proposed by \citet{schu68} for the study of the Hildas, extended to more variables to integrate as in
\citet{tm96} and \citet{mima04} but with a resonant condition like in \citet{go05}.

As we know, the disturbing function of a TBR is a second order function of the planetary masses,
which means  the calculation of the double integral (\ref{doble}) cannot be
done over the perturbing function evaluated at the unperturbed heliocentric positions. This can be shown considering that
$R = R_{01} + R_{02}$ and as there is no commensurability between the particle and the planet $P_1$ nor between
the particle and the planet $P_2$, the mean of  $R_{01}$ and  $R_{02}$ become independent of $\sigma$,
 then they do not contribute to $\mathfrak{R}(\sigma)$.
To properly evaluate the integral it is necessary to
take into account their mutual perturbations in the position vectors   $\vec{r_i}$.
Two body mean motion resonances are a simpler case because being a first order
perturbation in the planetary masses the position vectors
can be substituted by the Keplerian, non perturbed positions.

In previous works \citep{mhp98,nm99} analytical methods were developed in order to obtain
 the solution of an asteroid in TBR with Jupiter and Saturn. In this paper we are mainly interested not in obtaining
 the solution but in estimate the comparative strength of thousands of resonances with all the planets along all the Solar System.
Faced to this problem, in order to estimate the behavior of $\mathfrak{R}(\sigma)$
we adopt the following scheme for computing the double integral of Eq. (\ref{doble}):
\begin{equation}
 R(\lambda_0,\lambda_1,\lambda_2)  \simeq     R_u + \Delta R
\end{equation}
where $R_u$ stands from $R$ calculated at the unperturbed positions of the three bodies
and $\Delta R$ stands from the variation in $R_u$ generated by the perturbed (not Keplerian) displacements of the three bodies
in a  small interval $\Delta t$.
More clearly, given any set of the three position vectors $\vec{r_i}$ satisfying Eq. (\ref{sigma}) we compute the mutual perturbations
of the three bodies
and calculate the $\Delta\vec{r_i}$ that they generate in a small interval $\Delta t$ and the $\Delta R$ associated.
This scheme is equivalent to evaluate the integral over the infinitesimal trajectory
the system follows due to the mutual perturbations when released at all possible unperturbed positions that verify Eq. (\ref{sigma}).
We have then
\begin{eqnarray}
\label{R}
    R_u &=&   R_{01} +   R_{02} \\
    \Delta R &=& \Delta R_{01} + \Delta R_{02}
\end{eqnarray}
where $R_{01}$ and $R_{02}$ refer to the disturbing functions evaluated at the unperturbed positions and
$\Delta R_{01}$ and $\Delta R_{02}$ refer to the variations due to displacements caused by the mutual perturbations:

\begin{equation}
\label{dr01}
     \Delta R_{01} = \nabla_0 R_{01} \Delta \vec{r_0} +  \nabla_1 R_{01} \Delta \vec{r_1}
\end{equation}

\begin{equation}
\label{dr02}
     \Delta R_{02} = \nabla_0 R_{02} \Delta \vec{r_0} +  \nabla_2 R_{02} \Delta \vec{r_2}
\end{equation}
where $\Delta \vec{r_i}$  refers to displacements with respect to the heliocentric Keplerian motion
and being
\begin{equation}
\label{gradiRij}
    \nabla_i R_{ij}= k^{2} m_{j} (\frac{ \vec{r_j} -  \vec{r_i} }{r_{ij}^3}  - \frac{\vec{r_j}}{r_j^3})
\end{equation}
\begin{equation}
\label{gradjRij}
    \nabla_j R_{ij}= k^{2} m_{j} (\frac{ \vec{r_i} -  \vec{r_j} }{r_{ij}^3}  - \frac{\vec{r_i}}{r_j^3} + 3(\vec{r_i}\vec{r_j})\frac{\vec{r_j}}{r_j^5})
\end{equation}
From the equations of motion we have:
\begin{equation}\label{dx0dt2}
    \ddot{\vec{\Delta r_0}} =  \nabla_0 R_{01} +  \nabla_0 R_{02}
\end{equation}
\begin{equation}\label{dx1dt2}
    \ddot{\vec{\Delta r_1}} =  \nabla_1 R_{12}
\end{equation}
\begin{equation}\label{dx2dt2}
    \ddot{\vec{\Delta r_2}} =  \nabla_2 R_{21}
\end{equation}
Integrating twice we obtain the displacements with respect to the Keplerian motion:
\begin{equation}\label{dx0}
    \vec{\Delta r_0} \simeq  (\nabla_0 R_{01} +  \nabla_0 R_{02}) \frac{(\Delta t)^2}{2}
\end{equation}
\begin{equation}\label{dx1}
    \vec{\Delta r_1}  \simeq   \nabla_1 R_{12} \frac{(\Delta t)^2}{2}
\end{equation}
\begin{equation}\label{dx1}
    \vec{\Delta r_2} \simeq   \nabla_2 R_{21} \frac{(\Delta t)^2}{2}
\end{equation}
As the integral of $R_{u} = R_{01} + R_{02}$ becomes independent of $\sigma$,
we are only interested
in computing the function $\rho(\sigma)$ defined by
\begin{equation}\label{suma}
\rho(\sigma) = \frac{1}{4\pi^2}\int_{0}^{2\pi}d\lambda_1\int_{0}^{2\pi}\Delta R d\lambda_2
\end{equation}
always satisfying Eq. (\ref{sigma}), being
\begin{equation}\label{rfin0}
\Delta R= \frac{(\Delta t)^2}{2}
\left( (\nabla_0 R_{01})^2+(\nabla_0 R_{02})^2+2\nabla_0 R_{01}\nabla_0 R_{02}  +
  \nabla_1 R_{01} \nabla_1 R_{12}    +   \nabla_2 R_{02} \nabla_2 R_{21} \right)
\end{equation}
It is possible to show that the first two terms are independent of $\sigma$ then
\begin{equation}\label{rfin}
\rho(\sigma)= \frac{(\Delta t)^2}{2}
\frac{1}{4\pi^2}\int_{0}^{2\pi}d\lambda_1\int_{0}^{2\pi}
\left( 2\nabla_0 R_{01}\nabla_0 R_{02}  +
  \nabla_1 R_{01} \nabla_1 R_{12}    +   \nabla_2 R_{02} \nabla_2 R_{21} \right)d\lambda_2
\end{equation}
Note that $\rho(\sigma)\propto m_1 m_2$ while in the case of two body resonances the disturbing function is proportional to only one planetary mass
making TBRs much weaker than two body resonances.
We identify $\Delta t$ with the permanence time in each element of the phase space $(\Delta \lambda_0, \Delta \lambda_1, \Delta \lambda_2)$.
If the double integral is computed dividing the dominium in $N$ equal steps in $\lambda_1$ and
$N$ equal steps in $\lambda_2$ we can calculate the mean elapsed time  $\Delta t$ in the element of phase space
as
\begin{equation}\label{dt}
    \Delta t = \frac{\sqrt[3]{T_0 T_1 T_2}}{N}
\end{equation}
where $T_i$ are the orbital periods. Then
\begin{equation}\label{dtcua}
    \Delta t ^{2} = \frac{4\pi^{2} a_0 a_1 a_2}{k^2 N^2}
\end{equation}

The above algorithm  is independent of the pair of independent variables $\lambda_i$ used in the integral
and is independent of the order in which the double integral is evaluated. Taking $N$ equal for all resonances its
actual value is irrelevant; in our codes we use an arbitrary value $N=180$
and we have divided $\rho(\sigma)$ by  $k^4 m_{Jupiter}^2$ for convenience.
Considering $\sigma$ as a constant parameter we calculate the integral (\ref{rfin}) for a set of values of $\sigma$ between $(0,2\pi)$
and we obtain numerically $\rho(\sigma)$.
As defined above,  $\rho(\sigma)$ carries an arbitrary multiplicative constant and is dimensionally equivalent to a disturbing function.

\subsection{Shapes of $\rho(\sigma)$ and asymmetric equilibrium points}
\label{shape}

We must stress that the function $\rho(\sigma)$ is not the resonant disturbing function $\mathfrak{R}(\sigma)$, but closely related to it. While $\mathfrak{R}$ is a mean ideally computed along  the actual trajectories of the three bodies,
$\rho$ is computed along infinitesimal perturbed trajectories around the Keplerian positions of the three bodies.
Anyway, a strong dependence
of $\rho$ with $\sigma$ is indicative of a strong resonance. On the other hand, if the critical angle $\sigma$ does not affect  $\rho$ it will be indicative of a weak resonance.
Also, an extreme of  $\rho(\sigma)$ at some $\sigma$ means that
for that critical angle the perturbations have an extreme, that means, there is an
equilibrium point, but
we cannot apply stability criteria deduced for $\mathfrak{R}$ with  $\rho(\sigma)$ because they are not the same function, in particular
we cannot deduce wether the equilibrium points are stable or unstable.

For several resonances we confronted the shape of $\rho(\sigma)$ with numerical integrations of fictitious particles and with the analytical results by \citet{nm99}.
For low eccentricity orbits the most common case is a cosine or sine function for
 $\rho(\sigma)$ with equilibrium points at $0^{\circ}$ and $180^{\circ}$
degrees as in the theory by \citet{nm99}, but for some cases, approximately one third of the resonances studied, minima of  $\rho(\sigma)$  not necessarily mean stable nor maxima mean unstable.
We also found very common that for medium to high eccentricity orbits equilibrium points appear at asymmetric
positions $\sigma \neq 0^{\circ}, 180^{\circ}$.
We analyzed their existence and stability with numerical integrations of fictitious particles for several cases
using EVORB \citep{fgb02} as we show below, confirming in general their existence but with  compromised stability by
several perturbations we did not take into account in our algorithm.

\begin{figure}[h]
\resizebox{14cm}{!}{\includegraphics{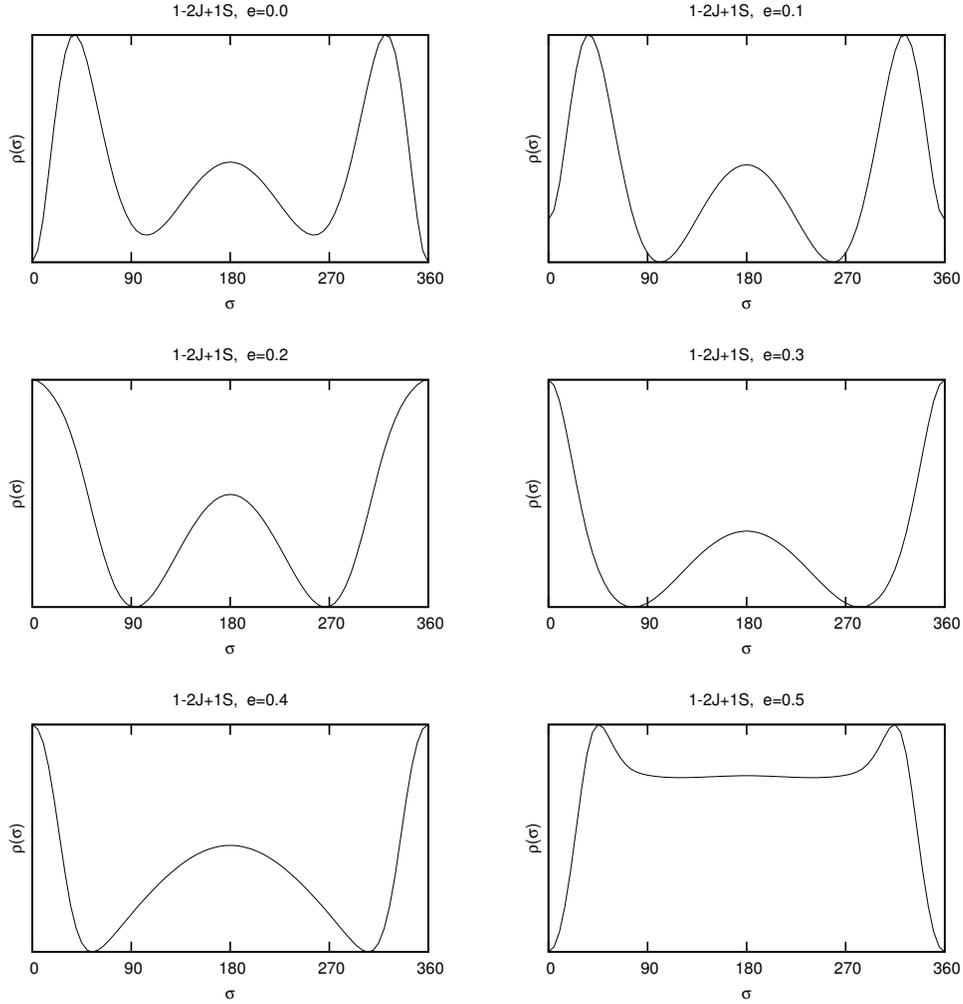}}
\caption{Evolution of the shape of $\rho(\sigma)$ for growing eccentricities for the zero order resonance 1-2J+1S at $a=3.8045$ au.
Numerical integrations of test particles show that
the asymmetric librations around $\sigma \sim 90^{\circ}$ and  $\sigma \sim 270^{\circ}$ are stable
and also exist large amplitude librations around $\sigma=180^{\circ}$ that wrap both asymmetric libration centers.
The librations around $\sigma=0^{\circ}$ are unstable.}
\label{order0}
\end{figure}

\begin{figure}[h]
\resizebox{14cm}{!}{\includegraphics{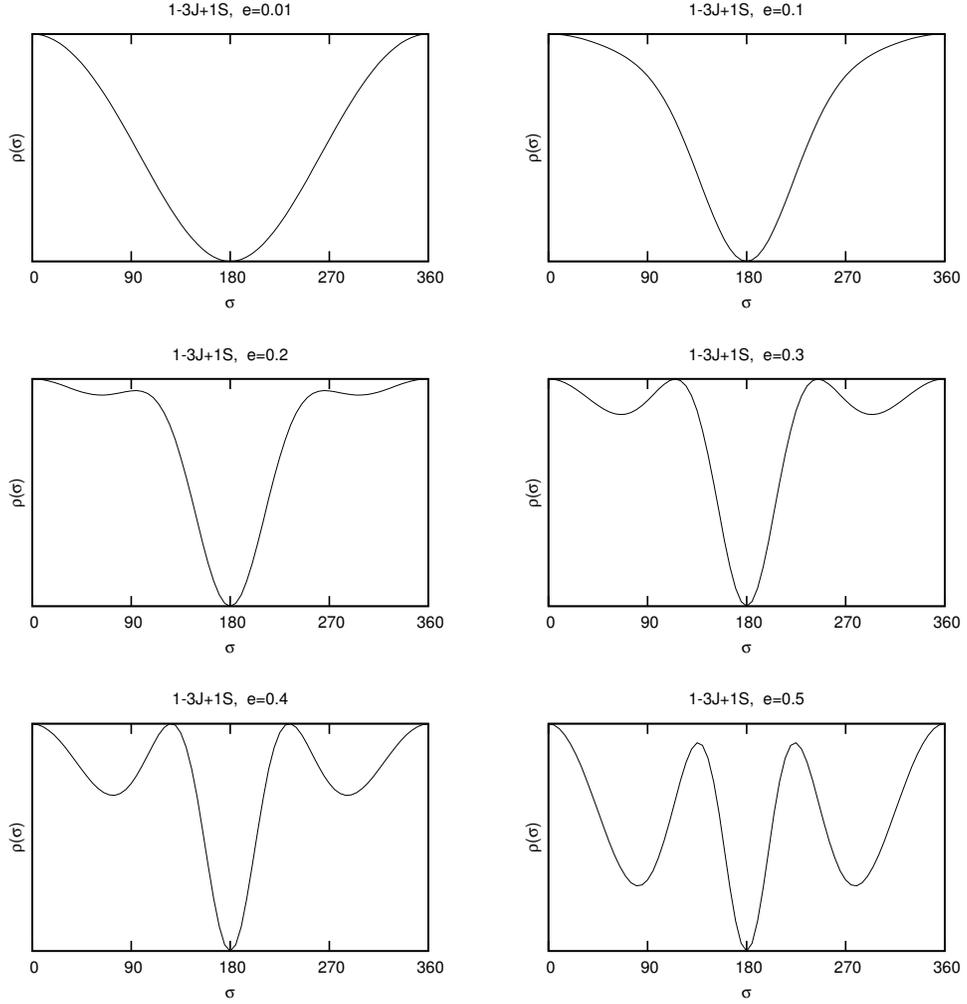}}
\caption{Evolution of the shape  of $\rho(\sigma)$  for growing eccentricities for the first order resonance 1-3J+1S at $a=2.7518$ au, the resonance where is evolving 485 Genua.
For large eccentricities asymmetric equilibrium points appear, which are stable according to numerical integrations of test particles.
There are also large amplitude librations around $\sigma=180^{\circ}$ that wrap both asymmetric libration centers.}
\label{genua}
\end{figure}

\begin{figure}[h]
\resizebox{12cm}{!}{\includegraphics{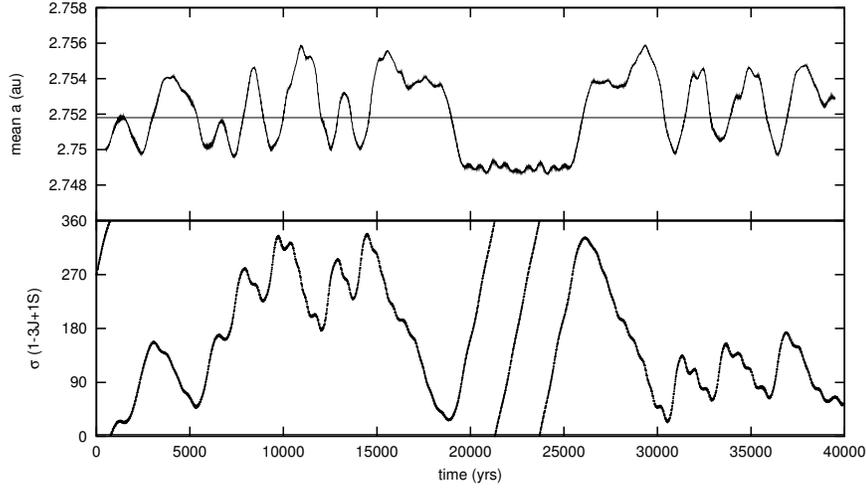}}
\caption{A fictitious particle with $0.3<e<0.4$ and $i < 4^{\circ}$ evolving in the real planetary system with temporary
captures in the asymmetric equilibrium points of the resonance 1-3J+1S. In top panel
the mean semimajor axis calculated as the mean value in a running window of 1000 yrs. The horizontal line indicates the nominal resonance.
Bottom panel: the time
evolution of the critical angle $\sigma = \lambda - 3 \lambda_J + \lambda_S + \varpi$.}
\label{asym}
\end{figure}

\begin{figure}[h]
\resizebox{12cm}{!}{\includegraphics{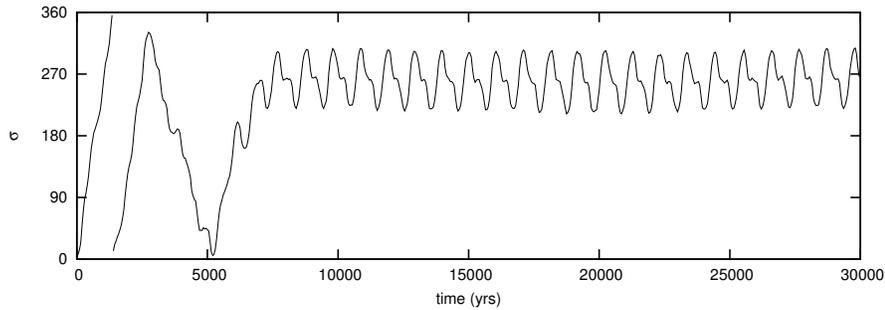}}
\caption{The critical angle for a fictitious particle with $e\sim 0.5$ and $i=0^{\circ}$
captured in an asymmetric equilibrium point
 in the
 the resonance 1-3J+1S assuming Jupiter and Saturn in circular coplanar orbits.}
\label{asymcircular}
\end{figure}

For example, in Figs. \ref{order0} and \ref{genua} we show the transformations of $\rho(\sigma)$ for increasing eccentricities for two resonances.
The equilibrium point at $\sigma=0^{\circ}$  in low eccentricity regime showed in Fig. \ref{order0} for the resonance
1-2J+1S is unstable, but on the other hand
the librations around the asymmetric centers at $\sigma \sim 90^{\circ}, 270^{\circ}$ are stable as our numerical integrations of fictitious particles confirm.
In the next case, referred to the resonance 1-3J+1S where inhabits 485 Genua,
for low eccentricity orbits our $\rho(\sigma)$ is very similar to the resonant disturbing function obtained by \citet{nm99}
but for $e>0.2$
equilibrium points appear at $90^{\circ}$ and $270^{\circ}$ which
we confirmed by numerical integrations of fictitious particles. Figure \ref{asym}
shows temporary captures of a fictitious particle evolving in the real planetary system in these asymmetric libration centers. In numerical integrations considering
only two planets in circular coplanar orbits these asymmetric librations are considerably more stable as showed in Fig. \ref{asymcircular}
where only Jupiter and Saturn in coplanar circular orbits were considered as perturbers.
The  librations around $\sigma=180^{\circ}$ for these two resonances are in fact trajectories that wrap the asymmetric librations points.
A careful reader will find a temporary libration around  $\sigma \sim 90^{\circ}$ for 485 Genua in Fig. 2 of \citet{nm98} at $t\sim 75000$ years. Analyzing several resonances we conclude that asymmetric librations are very common and
we can say that in general  $\rho(\sigma)$ is a good indicator to the existence of equilibrium points but their stability needs to be studied case by case.
Nevertheless, the asymmetric librations are easily destroyed by the perturbations due to the other non resonant planets, then its detection in
real asteroids could be difficult.

\subsection{Strengths   $\Delta\rho$ of the three body resonances}
\label{strength}

\begin{figure}[h]
\resizebox{12cm}{!}{\includegraphics{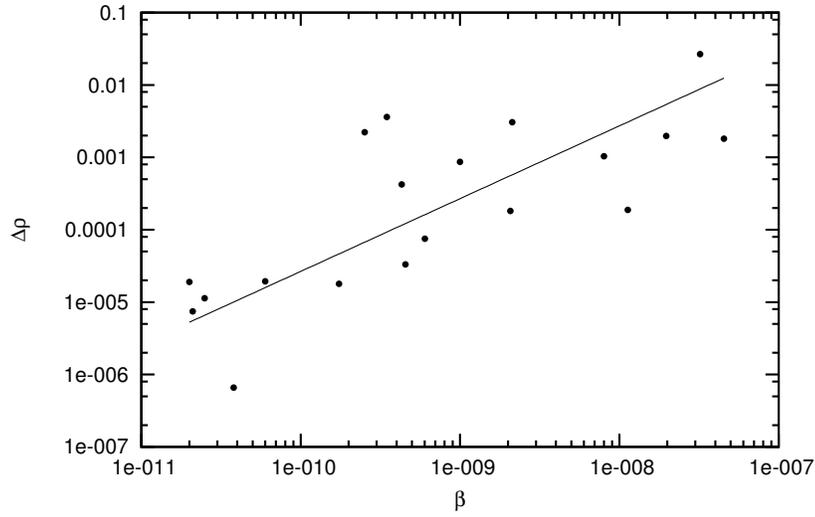}}
\caption{The semiamplitude $\Delta \rho$ calculated with
Eq. (\ref{rfin})
assuming $e=0.1, i=0^{\circ}$
as a
function of  the semiamplitude $\beta$ deduced from \citet{nm99} for the 19 resonances  of their Table I.
The curve fitting corresponds to  $\Delta \rho \sim  3\times 10^{5} \beta$.
}
\label{betadr}
\end{figure}

In order to obtain a relationship between $\rho$ and the disturbing function  $\mathfrak{R}$
we generated Fig.  \ref{betadr} comparing the semiamplitude $\Delta \rho =(\rho_{max}-\rho_{min})/2$ with the semiamplitude $\beta$ calculated by \citet{nm99} for the 19 resonances of their Table I.
This is a  very difficult comparison because
our calculations correspond to circular planetary orbits while \citet{nm99}
considered eccentric and time evolving orbits for Jupiter and Saturn which probably is one of the sources of the evident scatter in the figure.
The curve fitting indicated in the figure corresponds to $\Delta \rho \sim  3\times 10^{5} \beta$, but due to the errors associated
 we can only say that $\beta$ and $\Delta \rho$ are roughly proportional.

\begin{figure}[h]
\resizebox{12cm}{!}{\includegraphics{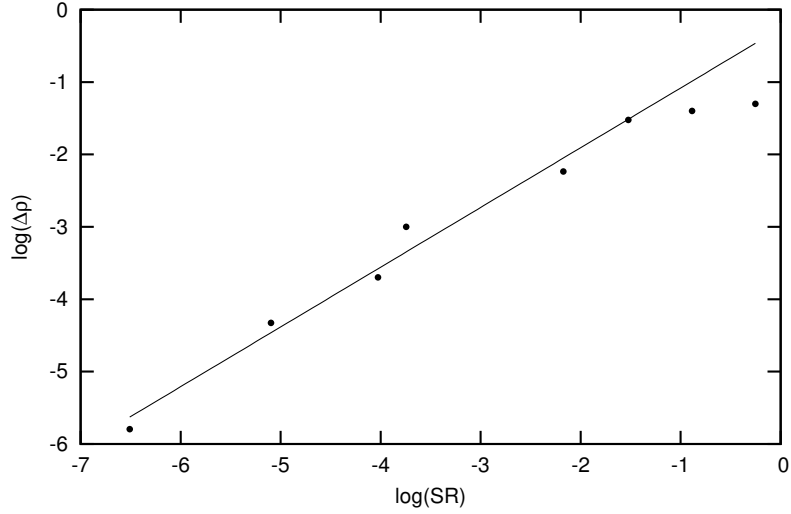}}
\caption{$\Delta \rho $ in logarithmic scale calculated with Eq. (\ref{rfin2})  for two body resonances versus the strength SR in logarithmic scale according to
\citet{ga06} for 9 arbitrary two body resonances assuming a circular orbit for the planet and
$e=0.1, i=0^{\circ}$ for the particle. The curve fitting corresponds to $\Delta \rho \propto \mathrm{SR}^{0.8}$. }
\label{srdr}
\end{figure}

We can devise another test for our $\Delta \rho$ applying it to the case of two body resonances. Following an analogue reasoning as devised here for TBRs
it is possible to show that in the case of a two body resonance  $\rho(\sigma)$  becomes
\begin{equation}\label{rfin2}
\rho(\sigma)= \frac{(\Delta t)^2}{2}
\frac{1}{2\pi}\int_{0}^{2\pi} (\nabla_0 R_{01})^2 d\lambda_1
\end{equation}
We calculated  the semiamplitude $\Delta \rho$ for some two body resonances and compared with the strength, SR, given by \citet{ga06} which in this case was
calculated assuming  a circular orbit for the planet.
The result indicating a more clear relationship $\Delta \rho \propto \mathrm{SR}^{0.8}$  is showed in Fig. \ref{srdr}.
With these two tests we can conclude that   $\Delta \rho$ is nearly proportional to the semiamplitude of the resonant disturbing function
and in spite of the limitations of the method
can be considered a rough measure of the resonance's strength.
We will use this quantity for studying TBRs, since it is easy to compute  and a good compromise between analytical understanding and accuracy on one hand, and effective exploration of a vast parameter space, on the other.

\begin{figure}[h]
\resizebox{12cm}{!}{\includegraphics{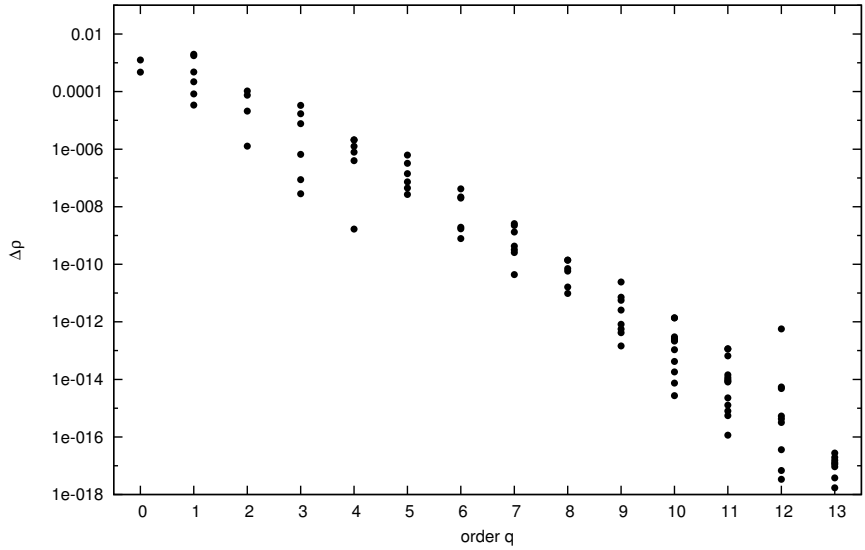}}
\caption{ $\Delta \rho$ versus order $q$  for  TBRs of order $ q \leq 13$ and $p \leq 20$ with Jupiter and Saturn from 2 to 2.4 au, calculated assuming $e=0.1$ and $i=0^{\circ}$.
It is approximately verified that $\lg (\Delta \rho) \propto - q$. }
\label{srjs}
\end{figure}

\begin{figure}[h]
\resizebox{12cm}{!}{\includegraphics{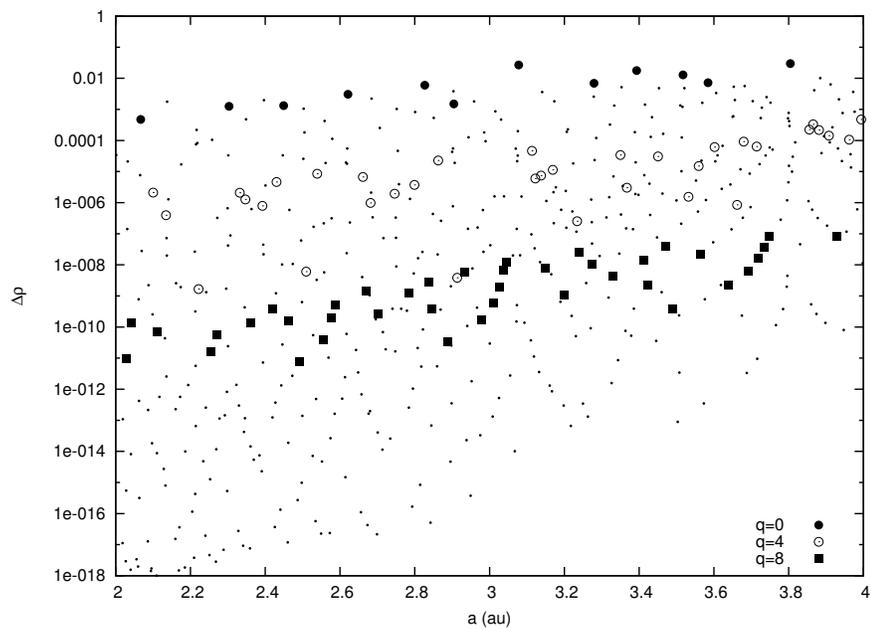}}
\caption{ $\Delta \rho$ versus semimajor axis for TBRs of order $ q \leq 13$ and $ p \leq 20$ with Jupiter and Saturn from 2 to 4 au, calculated assuming $e=0.1$ and $i=0^{\circ}$ where resonances of order 0, 4 and 8 are indicated with different symbols for comparison. Zero order resonances and resonances located closer to the planets tend to be stronger.}
\label{srjs2}
\end{figure}

To analyze the properties of  $\Delta \rho$
we calculated it for 489 TBRs with Jupiter and Saturn with order $q\leq 13$ and $p \leq 20$
located between 2 and 4 au summarizing the results in  Figs. \ref{srjs} and \ref{srjs2}.
From Fig. \ref{srjs} it is evident
an exponential decay
 of  $\Delta \rho$ with the order $q$ as theoretical models predict and from  Fig. \ref{srjs2} it is clear also a dependence with the proximity
to the planets;
resonances closer to the planets tend to be stronger as pointed out by \citet{nm98}.
We found an analogue behavior, approximately symmetric,  for the external TBRs  with Uranus and Neptune in the TNR.
Note in Fig. \ref{srjs2} the pattern at $a\sim 3.8$ au related to the superposition of two two-body resonances: 1-4S at $a=3.7915$ au and 5-8J at $a=3.8021$ au
which added are very close to the TBR
3-4J-2S at $a=3.8002$ au.
As the resonant condition (\ref{sigma}) can also be written $(k_0+j)\dot{\lambda_0}+k_1\dot{\lambda_1}= j\dot{\lambda_0}-k_2\dot{\lambda_2} =\nu$,
being $j$ an arbitrary integer, if $a_0$ is such that for some $j$ it is verified $\nu\simeq 0$, then the TBR will be the resulting of the superposition of two two-body resonances and a particular arrangement of TBRs appears near $a_0$.

\begin{figure}[h]
\resizebox{12cm}{!}{\includegraphics{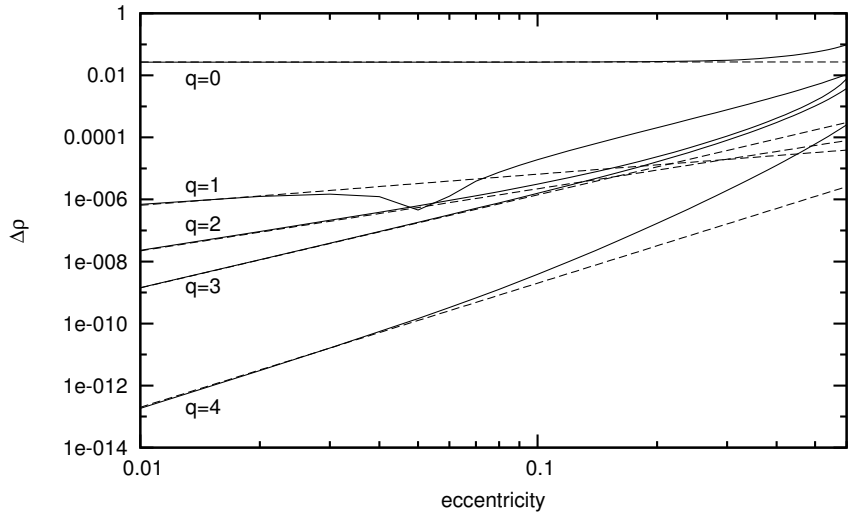}}
\caption{Continuous lines: $\Delta \rho$ as function of the asteroid's eccentricity for zero ($q=0$, 1-3J+2S at 3.0777 au), first  ($q=1$, 1-4J+4S at 2.9067 au), second  ($q=2$, 2-7J+7S at 3.1773 au), third  ($q=3$, 1-5J+7S at 3.0854 au) and fourth ($q=4$, 1-6J+9S at 2.9134 au) order
resonances assuming  $i=0^{\circ}$. At low eccentricities these curves correspond approximately with $\Delta \rho \propto e^{q}$
whose representations are showed with dashed lines. The depression of the curve corresponding to the
resonance 1-4J+4S at $e\sim 0.05$ is related to the appearance and disappearance of asymmetric equilibrium points.}
\label{drzero}
\end{figure}

\begin{figure}[h]
\resizebox{12cm}{!}{\includegraphics{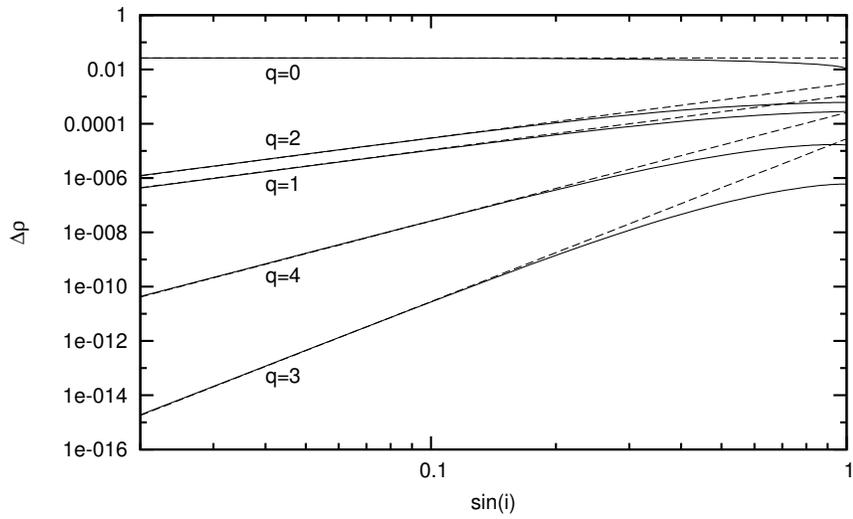}}
\caption{Continuous lines: $\Delta \rho$ as function of the asteroid's inclination for the same
resonances of Fig. \ref{drzero} but  assuming  $e=0$ and $\omega=0^{\circ}$.
For low inclinations these curves correspond approximately with  $\Delta \rho \propto (\sin i)^{z}$ where $z=q$ for even order and $z=2q$ for odd order resonances.
Dashed lines: curve fitting to $(\sin i)^{q}$ for even order resonances and to $(\sin i)^{2q}$ for odd order resonances.}
\label{drten}
\end{figure}

The dependence with the eccentricity and the order of the resonance is showed in Fig. \ref{drzero} for orbits with $i=0^{\circ}$ and
the dependence with the inclination and the order of the resonance is showed in Fig.
\ref{drten}  for $e=0$ respectively.
With these and others plots going up to $q=5$ we deduced by curve fitting that for zero inclination orbits $\Delta \rho \propto e^{q}$, which is coherent with
  theoretical models for coplanar orbits that predict the lower order terms in the disturbing function are factorized by
   $e^{q}$.
Moreover, we find that for circular orbits
$\Delta \rho \propto (\sin i)^{q}$  for even order resonances and
$\Delta \rho \propto (\sin i)^{2q}$ for odd order resonances.
That means, our algorithm predicts that the lower order terms in the disturbing function for circular orbits would be factorized by
$(\sin i)^{q}$ for $q$ even and by $(\sin i)^{2q}$ for $q$ odd.
We have no knowledge of theoretical predictions about this type of dependence of the
strength on the orbital inclination in the case of TBRs but we know that two body resonances have also an asymmetric behavior between eccentricity and inclination.

It is interesting to note that two body resonances and TBRs share the same property: their strengths are proportional to $e^{q}$ which means that quasi circular resonant orbits are weakly bounded except for zero order resonances which are
the strongest ones, they are almost independent of the eccentricity and have some strength
even for zero eccentricity orbits. But there is a fundamental difference: zero order two body resonances are confined to co-orbital motions (trojans, for example)
while zero order TBRs occur in a wide variety of orbital configurations of the three bodies.
Then,  in a migration scenario in a protoplanetary disc, bodies in quasi circular orbits could be trapped in zero order TBRs since these resonances  have some associated strength \citep{qu11}.

\section{Atlas of the three body resonances in the Solar System}
\label{atlas}

\begin{figure}[h]
\resizebox{14cm}{!}{\includegraphics{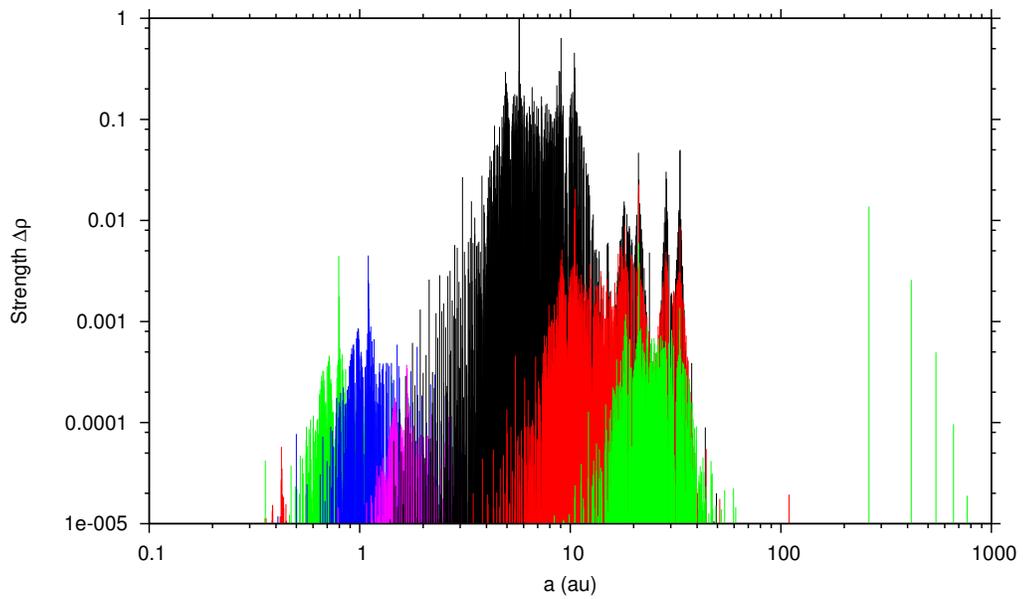}}
\caption{Global view of the atlas of TBRs from 0.1 to 1000 au assuming $e=0.15, i=6^{\circ}, \omega=60^{\circ}$.
Color version: resonances which its most interior planet is Mercury, Venus, Earth, Mars, Jupiter, Saturn or Uranus are showed in red,
green, blue, pink, black, red or green respectively.
 Note the strong resonances involving Uranus and Neptune at $a>250$ au.}
\label{global}
\end{figure}

\begin{figure}[h]
\resizebox{14cm}{!}{\includegraphics{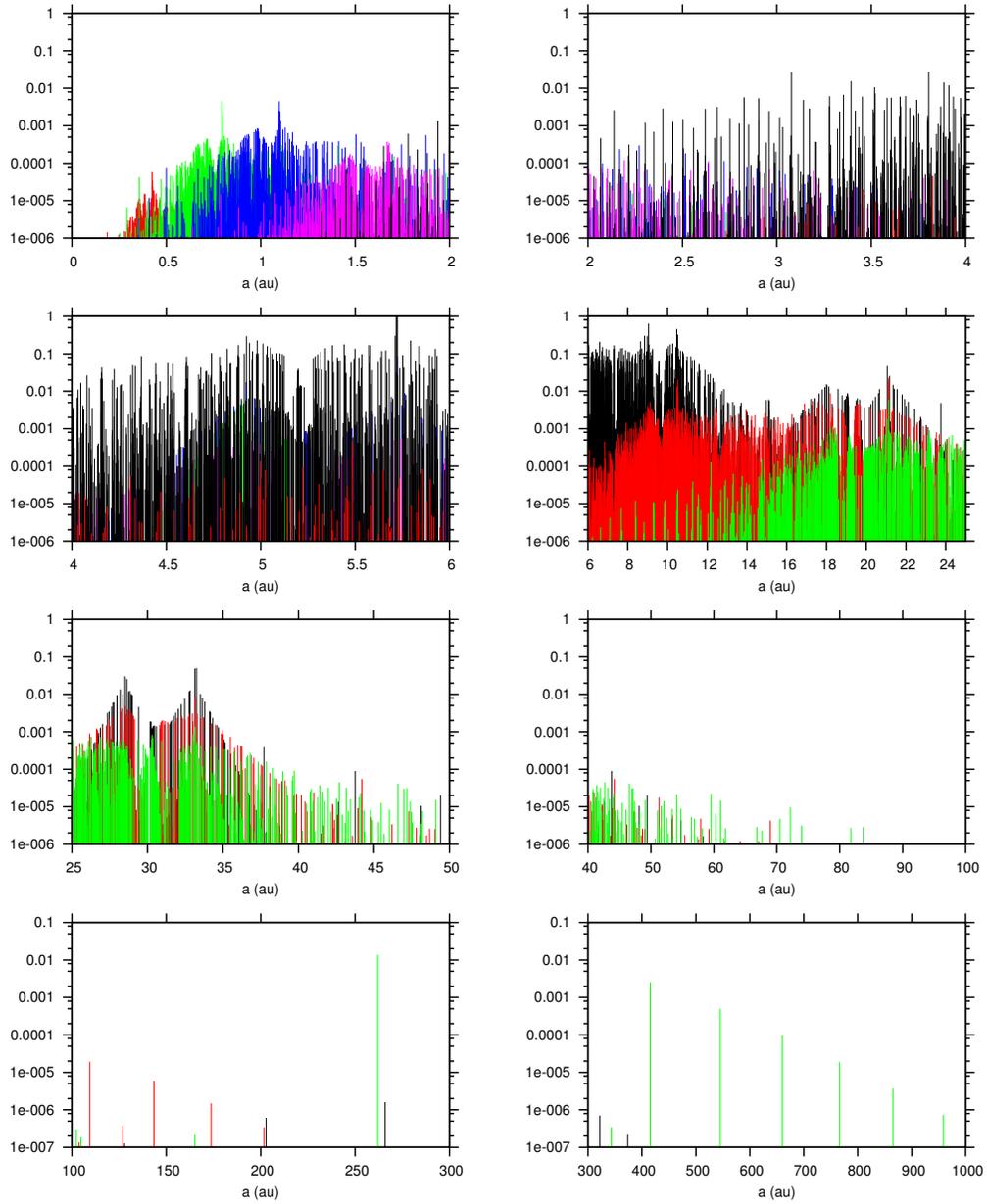}}
\caption{Same as Fig. \ref{global} with linear scale in semimajor axis. Qualitative comparison with Fig. 7 from \citet{ga06} can be done.}
\label{plot8}
\end{figure}

For each pair of perturbing planets assumed in circular coplanar orbits, the strength  $\Delta \rho$ depends on the semimajor axis of the TBR defined by $(k_0,k_1,k_2)$ and on
the orbital elements $(e,i,\omega)$ of the massless particle.
We computed $\Delta \rho$
for all TBRs with  $p \leq 20$
involving pairs of the 8 planets from 0 to 1000 au assuming a test particle with $e=0.15$, $i=6^{\circ}$,
which is typical of the  main asteroid belt and an arbitrary $\omega=60^{\circ}$.
The general view of the distribution of these 55814 resonances is showed in Figs. \ref{global} and \ref{plot8} which can be qualitatively compared with the
atlas of two body resonances presented in Fig. 7 of \citet{ga06}, but no quantitative comparison can be done because  our $\Delta \rho$
is not exactly the amplitude of the disturbing function of the TBRs and the scales are different.

In Tables \ref{strongest10}  and \ref{strongest1000} we present some of the  strongest resonances along the Solar System  up to 1000 au where resonances involving
Jupiter and Saturn dominate except in the TNR where the resonances involving Uranus and Neptune dominate instead.
Table  \ref{terrjov}  shows the  strongest resonances  involving
terrestrial planets with jovian planets and
Table \ref{onlyterr}  shows the  strongest resonances  involving
only terrestrial planets.

There is a very dense region of TBRs between 0.5 a 2 au mainly due to resonances involving Venus-Earth, Earth-Jupiter, Venus-Jupiter and Earth-Mars.
For example, the strong resonances at $a\sim 0.8$ and $a\sim 1.1$ au are
due to
8-7V+1J  and
7-6E-1J respectively.
 Between 2 and 4 au there are some known strong resonances involving Jupiter and Saturn immerse in a region of low density of resonances.
 Between 4 and 35 au there is a region with high density of strong resonances involving the jovian planets. From 35 to 200 au resonances are weaker with the exception of
 2+1S-3U and
 3+1S-3U
 at $a\sim 109.5$
 and  $a\sim 143.4$ au respectively. Finally, for $a>250$ au it appears the series of unexpectedly strong resonances
 1+1U-2N, 2+1U-2N, 3+1U-2N and so ones.

We remark that our method ignores the planetary eccentricities then we expect the effective strengths in the real Solar System will be greater.
Moreover, as our method ignores variations in longitudes of the perihelia and nodes, each resonance will be composed by a multiplet
that our method cannot discern.

\subsection{Density of TBRs along the Solar System}

\begin{figure}[h]
\resizebox{12cm}{!}{\includegraphics{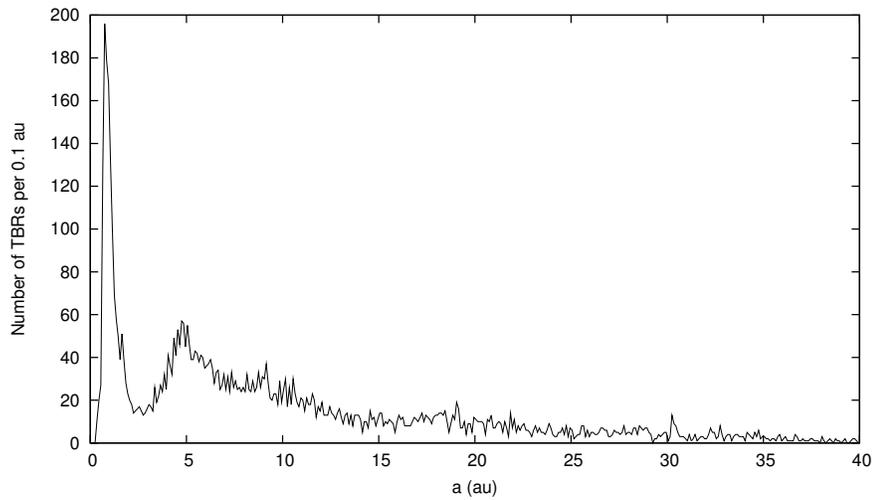}}
\caption{Number of TBRs with $\Delta \rho >10^{-5}$ per 0.1 au.  $\Delta \rho$ calculated assuming $e=0.15, i=6^{\circ}, \omega=60^{\circ}$. }
\label{density1}
\end{figure}

\begin{figure}[h]
\resizebox{12cm}{!}{\includegraphics{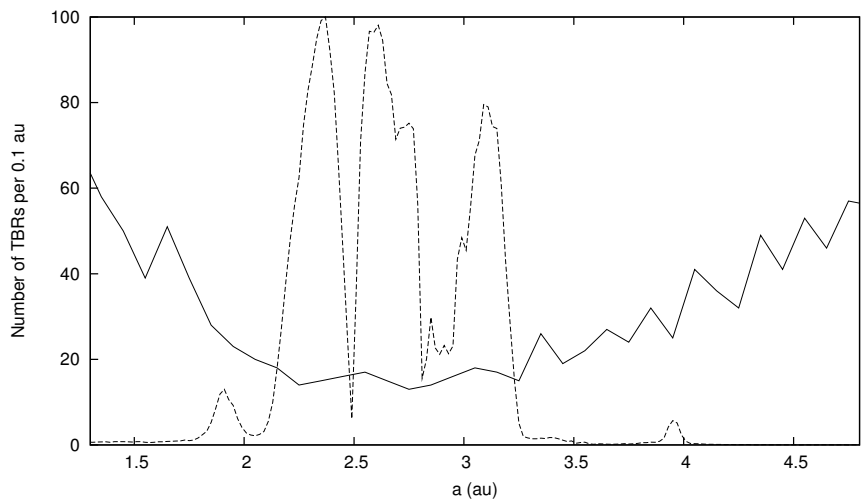}}
\caption{Detail of Fig. \ref{density1} including the normalized distribution of the osculating semimajor axis of the asteroids from ASTORB database
(dashed line). }
\label{density2}
\end{figure}

From the inspection of Table \ref{strongest10} we conclude that the strongest TBRs in the main asteroid belt have  $\Delta \rho \sim 10^{-3}$, then
to find how the dynamically relevant TBRs are distributed in the Solar System
 we have calculated the
 number of TBRs with $\Delta \rho > 10^{-5}$
 between
 intervals of 0.1 au
from 0 to 40 au showing the result in Fig. \ref{density1}.
As the superposition of TBRs is associated with chaotic dynamics \citep{mh97,mhp98,nm98,nm99,mn99},
a larger density implies a  more chaotic dynamics. Figure \ref{density1} can be considered as a global indicator of the chaos
generated by the TBRs in the Solar System, but to make an unequivocal diagnosis of chaos it is necessary to know the widths of the resonances
expressed in au,
point that our method in its present form cannot resolve for now.
The highest peak is between $0.7-1.0$ au and is produced mainly by
TBRs including Venus as the innermost planet and the Earth as the innermost planet. It is evident that the main asteroid belt is
located in the region with lowest density of TBRs between the planets.
To appreciate with more detail the situation in the asteroids' region  we plotted in Fig. \ref{density2}
a detail of Fig. \ref{density1} jointly with an histogram of the osculating semimajor axes of the asteroids taken from ASTORB database (\verb"ftp://ftp.lowell.edu/pub/elgb/astorb.html").
It is suggestive that the
stable population of asteroids is precisely located in the region with lowest density of TBRs in the Solar System.
At both sides of the main asteroid belt there is a clear increase in the density.

Considering only resonances with  $\Delta \rho > 10^{-5}$, at the left of the asteroid belt
between 1.5 and 2.0 au the most common relevant resonances involve Mars-Jupiter (47\%) and Earth-Jupiter (25\%),
inside the asteroid belt
between 2.0 and 3.3 au involve mostly Jupiter-Saturn (36\%), Mars-Jupiter (27\%) and Earth-Jupiter (21\%) and
at the right of the asteroid belt between 3.3 and 4.0 au the most common resonances involve Jupiter-Saturn (51\%) and Jupiter-Uranus (23\%).

\subsection{Signatures of TBRs in the main asteroid belt}

\begin{figure}[h]
\resizebox{12cm}{!}{\includegraphics{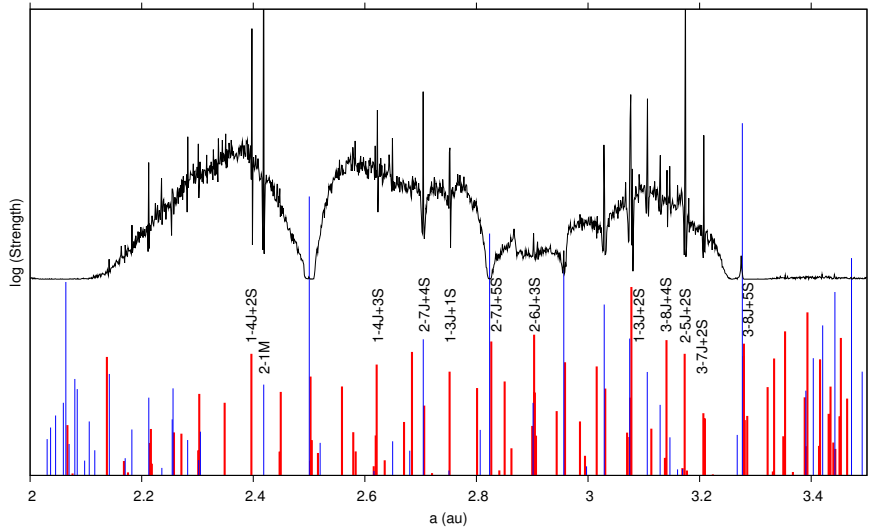}}
\caption{Strongest two body resonances (thin/blue lines) and TBRs (thick/red lines) with an histogram of
proper semimajor axes from AstDyS.
The strengths in logarithmic scale for both types of resonances were calculated assuming planets with circular orbits and the massless particle with $e=0.2, i=10^{\circ}, \omega=60^{\circ}$.
The sets of two body and TBRs are not in the same scale. Some of the strongest TBRs are labeled.}
\label{astdys}
\end{figure}

We can associate  details of the distribution of asteroids to two body and TBRs as showed in Fig. \ref{astdys}
 where it was plotted an histogram of the synthetic proper semimajor axes computed numerically given by AstDyS database (\verb"hamilton.dm.unipi.it/astdys/") using bins of 0.001 au. The peaks are due to concentrations
generated by two body and TBRs.
Note for example the strong peak at $a\simeq 2.419$ au due to the exterior resonance 2-1M \citep{ga07}.
By simple inspection of Fig. \ref{astdys} we find peaks in the histogram
where two body resonances are absent but TBRs are present like
1-4J+2S at $2.397$ au,
1-4J+3S at $2.622$ au,
1-3J+1S at $2.752$ au,
2-5J+2S at $3.173$ au and
3-7J+2S at $3.207$ au.
All these were already studied by  \citet{nm98} and \citet{ss13} found hundreds of asteroids evolving in all them except in the last one. A comparison with Fig. 1 from \citet{mn99} will also help to
identify the peaks in the histogram.
Our method indicates that the strongest TBR in the asteroid belt is 1-3J+2S (see also Table \ref{strongest10}) which is the second strongest
resonance according to \citet{nm99} and one of the three most populated resonances according to \citet{ss13}.

\subsection{Identifying a resonance: the case of 2009 SJ18}

\begin{figure}[h]
\resizebox{12cm}{!}{\includegraphics{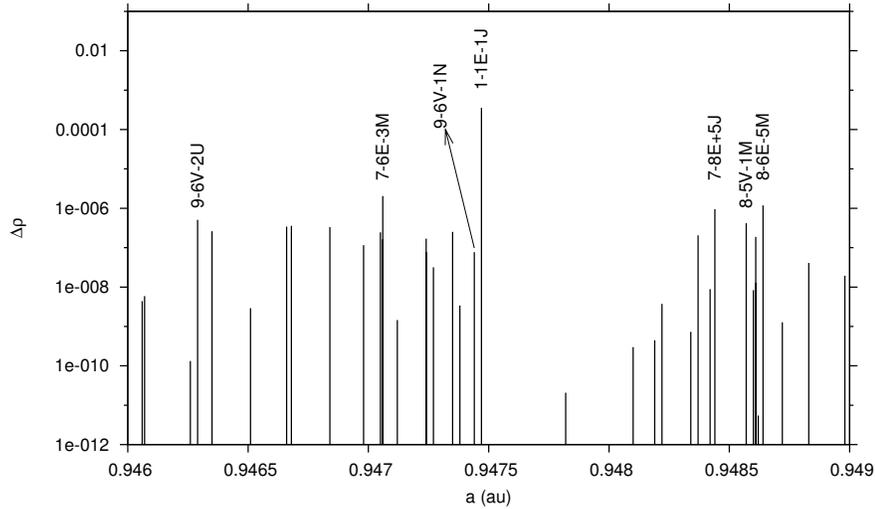}}
\caption{Strength $\Delta\rho$ calculated taking $e=0.11, i=13^{\circ}, \omega=173^{\circ}$ for all TBRs with $p\leq 20$ located near the
semimajor axis of 2009 SJ18.}
\label{2009sj18atlas}
\end{figure}

\begin{figure}[h]
\resizebox{12cm}{!}{\includegraphics{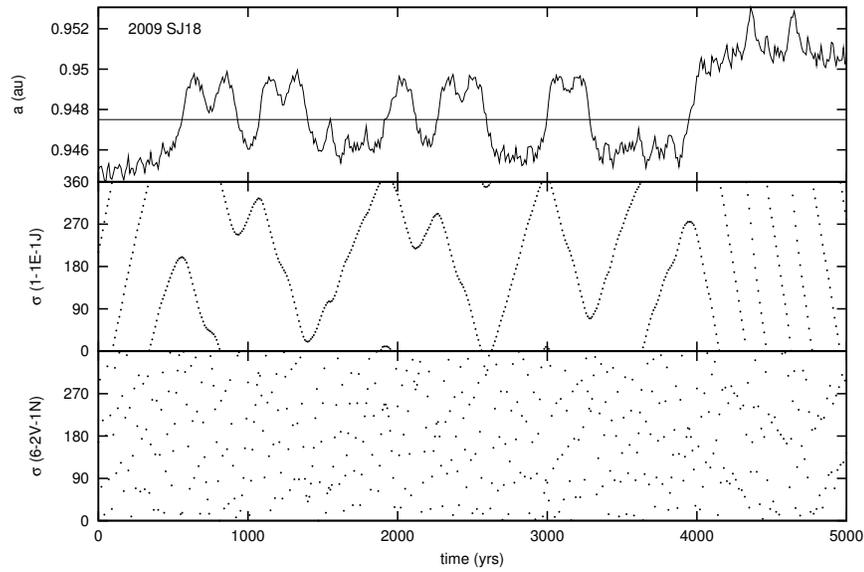}}
\caption{NEA 2009 SJ18 evolving in 1-1E-1J at $a=0.94747$ au with strength $\Delta\rho \simeq 3.6\times 10^{-4}$ showing librations around $\sigma = 180^{\circ}$ and transitory
librations around the asymmetric libration center at $\sigma \sim 270^{\circ}$ (mid panel). In the low panel
it is showed for comparison the critical angle related to the closest resonance to 1-1E-1J which is 9-6V-1N located at $a=0.94744$ au with strength $\Delta\rho \simeq 8\times 10^{-8}$.}
\label{2009sj18}
\end{figure}

We illustrate the use of the method with the case of the NEA 2009 SJ18.
A numerical integration of this object shows that its semimajor axis  oscillates between
0.945 and 0.950 au. Taking the present orbital elements $e,i,\omega$ (precise values are not necessary) for 2009 SJ18 we calculate $\Delta\rho$ for all TBRs with $p\leq 20$ located
 between these values of semimajor axis.
The strongest resonances are showed in Fig. \ref{2009sj18atlas} where  1-1E-1J dominates by two orders of magnitude with respect to its neighbors. Then, we calculate the time evolution of the
critical angle $\sigma = \lambda - \lambda_E -\lambda_J + \varpi$ and we plot $a(t)$ and $\sigma(t)$ at top and middle panels in Fig. \ref{2009sj18} respectively. In spite of being very close to Earth, it survives 4000 years captured in the resonance
1-1E-1J at $a=0.9475$ au.
The correlation between $a(t)$ and $\sigma(t)$ confirms the object is inside the resonance, but to discard a casual coincidence we also show at bottom panel in Fig. \ref{2009sj18} the time evolution of the
critical angle $\sigma = 9\lambda - 6\lambda_V -\lambda_N - 2\varpi$ corresponding to the closest resonance to 1-1E-1J   seen
in Fig. \ref{2009sj18atlas}. The circulation of this critical angle indicates that, in spite of this resonance being located very near to the value
of the asteroid's semimajor axis, the asteroid is not
 evolving under its influence, as we could have deduced from its negligible strength showed in
Fig. \ref{2009sj18atlas}. The critical angle corresponding to the resonance 7-6E-3M, the second strongest resonance in the interval analyzed,
also circulates.
This procedure consisting in calculating the strengths $\Delta\rho(e,i,w)$, being $(e,i,w)$ the orbital elements of the
asteroid we are studying, for all resonances with nominal location near the asteroid's semimajor axis, looking
for the strongest ones and calculating their critical angles, becomes very useful to identify the TBRs affecting the asteroid,
if there is any.
Other NEAs evolving in
resonances involving terrestrial planets can be detected  by a similar procedure.

\subsection{Some cases of TNOs and centaurs}

\begin{figure}[h]
\resizebox{12cm}{!}{\includegraphics{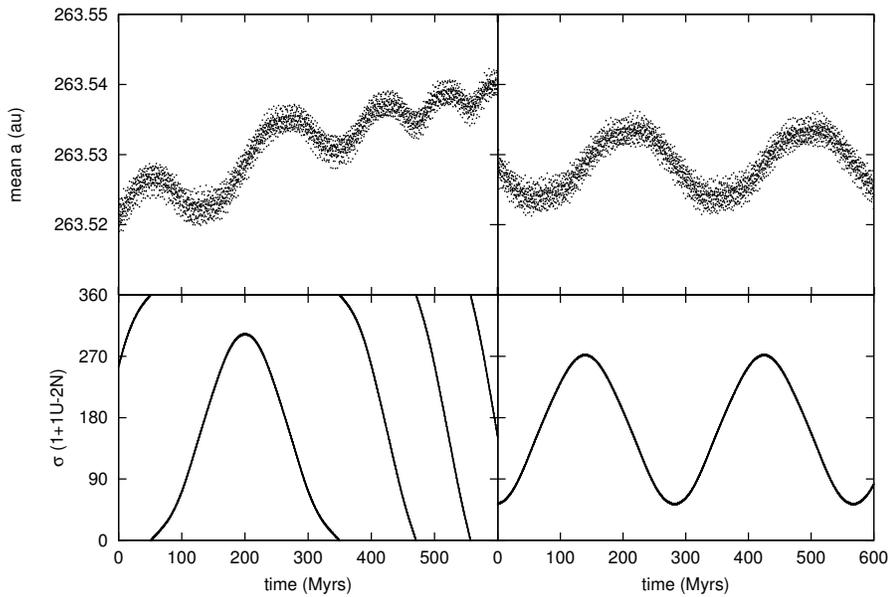}}
\caption{Two particles with an imposed migration in semimajor axis interacting with the resonance 1+1U-2N
in a fictitious planetary system composed only by the Sun and the planets Uranus and Neptune in coplanar quasi circular
orbits. Left panels: the particle is initially outside the resonance and
crossed the resonance due to the migration. Right panels: the particle is initially inside the resonance and resists the forced migration due to the resonance's strength.
The mean $a$ is calculated with a window of 1 Myrs.  The nominal position of the resonance for this particular planetary system is $a_0 = 264.93$ au but its
 actual position is shifted due to the motion of the perihelia that were ignored in Eq. (\ref{nominal}). All particles have $e=0.2$ and the resonance's strength is  $\Delta\rho \simeq 0.014$.}
\label{crossing}
\end{figure}

We call the attention to the  resonances 1+1U-2N, 2+1U-2N, 3+1U-2N and so ones in the far TNR.
They are unusually strong in a region where other TBRs  are several orders of magnitude weaker.
In order to appreciate the dynamical effects of these TBRs we numerically integrated a set of particles with
initial semimajor axes $a\leq a_{0}$ being $a_{0}$  the nominal location of the resonance 1+1U-2N
 and imposing a continuous perturbation so that the particles' semimajor axes slowly increase with time and eventually cross or get trapped into the resonance. We considered only the planets Uranus and Neptune in
 coplanar circular orbits
 in order to make the experiment under the hypotheses of our algorithm.
 The dynamical effect of the resonance in the semimajor axis is clearly seen in
 figure \ref{crossing} which shows the evolution of two particles, one that crosses the resonance and other that started inside and remains captured overcoming the forced migration with
the resonance's strength which in this experiment is  $\Delta\rho \simeq 0.014$.

\begin{figure}[h]
\resizebox{12cm}{!}{\includegraphics{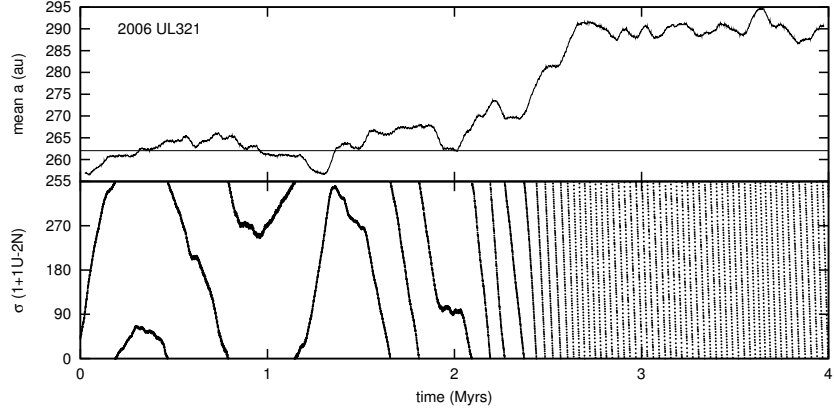}}
\caption{Scattered Disk Object 2006 UL321  evolving chaotically near  1+1U-2N
with resonance's strength $\Delta\rho \simeq 0.007$.  Mean $a$ calculated with a running window of 50000 yrs.}
\label{2006ul321}
\end{figure}

\begin{figure}[h]
\resizebox{12cm}{!}{\includegraphics{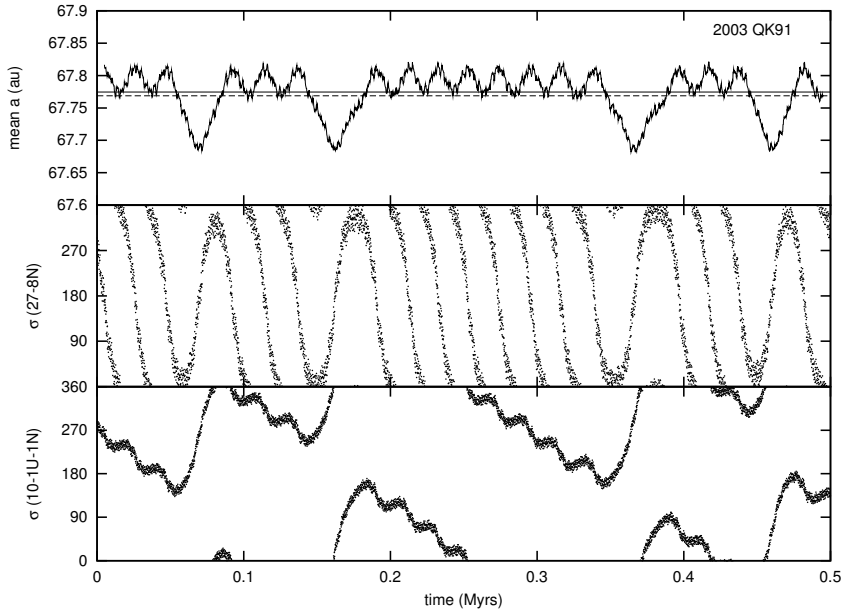}}
\caption{Evolution of 2003 QK91 dominated by the two-body resonance 27-8N (mid panel) and by the TBR 10-1U-1N (bottom panel)
with strength  $\Delta\rho \simeq 4\times 10^{-5}$. Mean $a$ computed using a running window of  10000 yrs. Dashed line indicates the nominal 27-8N and the continuous line indicates the nominal 10-1U-1N.}
\label{2003qk91}
\end{figure}

\begin{figure}[h]
\resizebox{12cm}{!}{\includegraphics{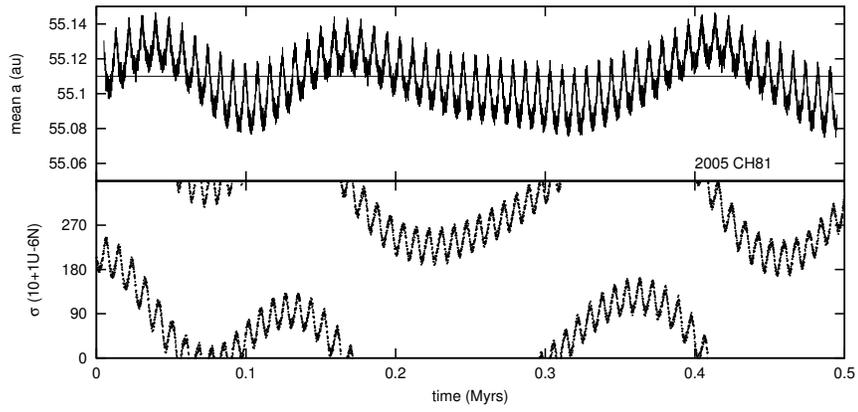}}
\caption{Evolution of 2005 CH81 near resonance 10+1U-6N. Mean $a$ computed using a runing window of  10000 yrs.
The nominal resonance is indicated with the line and its strength is $1.4\times 10^{-5}$.}
\label{2005ch81}
\end{figure}

Regarding the actual population of TNOs, 2006 UL321 has a very preliminary orbit determination, but taking their nominal elements from JPL (\verb"ssd.jpl.nasa.gov/sbdb_query.cgi") we obtained
an orbital evolution near 1+1U-2N with  $\Delta\rho \simeq 0.007$ showed in Fig. \ref{2006ul321}.
We performed numerical integrations of 87269 (2000 OO67) with clones and all them resulted evolving chaotically
near the resonance 3+1U-2N.
Other objects evolving in resonances with Uranus and Neptune are for example
Eris, which is evolving very near the resonance 10-1U-1N \citep{ga12} with strength $3\times10^{-5}$, 2003 QK91 which is also evolving in that resonance with similar strength
but with a strong influence of the two-body resonance 27-8N
as shown in Fig. \ref{2003qk91} and 2005 CH81 which is evolving   near the resonance 10+1U-6N
as shown in Fig. \ref{2005ch81}  with strength $1.4\times10^{-5}$.

\begin{figure}[h]
\resizebox{12cm}{!}{\includegraphics{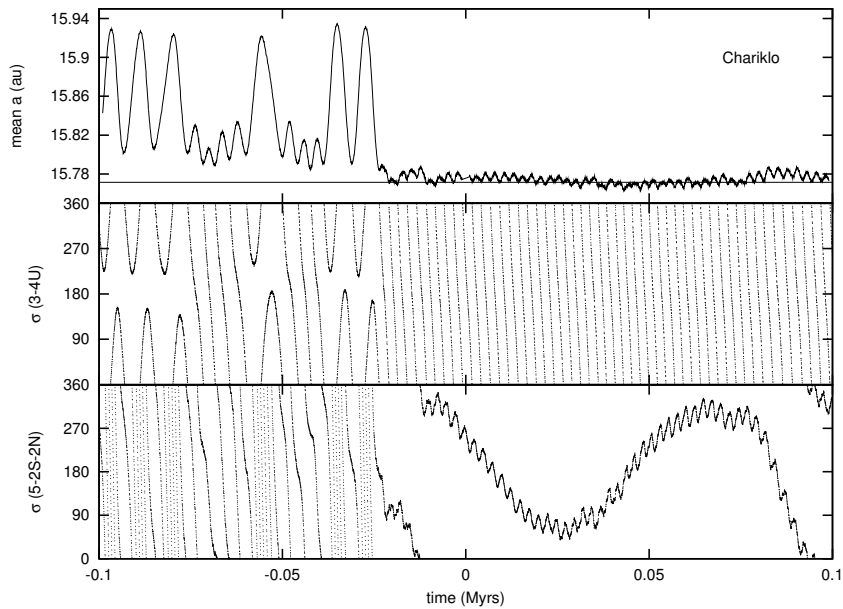}}
\caption{Centaur 10199 Chariklo in the past appears captured in the resonance 3-4U at $a=15.87$ au (mid panel) but at
$t\simeq -23800$ yrs its semimajor axis drops very close to the nominal value of the  TBR 5-2S-2N
at $a= 15.771$ au (bottom panel) with strength $\Delta\rho \simeq 4\times 10^{-4}$
being captured in this resonance since then. The horizontal line in top panel indicates
the location of the nominal TBR. Mean $a$ over 2000 yrs.}
\label{chariklo}
\end{figure}

A backwards integration shows the centaur 10199 Chariklo was captured
in the resonance 3-4U at a = 15.87 au,
 but at present is outside this resonance
 in a region where, according to its orbital elements, TBRs 5-2S-2N at 15.771 au, 9-1J-5U at 15.774 au and 7-4S+2U 15.776 au dominate
 with strengths $4\times10^{-4}$, $7\times10^{-4}$ and $9\times10^{-5}$ respectively.
 Figure \ref{chariklo} shows the critical angle for the resonance 3-4U in mid panel
 and the critical angle for the TBR 5-2S-2N in the low panel. The critical angle for 9-1J-5U circulates and the critical angle for 7-4S+2U
 slowly circulates.

Finally, we want to stress that in our numerical integrations of real and fictitious objects in the outer Solar System and in the TNR, when looking for TBRs
we found very common also the capture in high order two body resonances with Neptune or Uranus as was the case of Chariklo and 2003 QK91. The interaction
between TBRs and high order two body resonances seems to be a common dynamical situation.

\section{Conclusions}
\label{conc}

We defined a function $\rho(\sigma)$ related to the resonant disturbing function which allows us to estimate
 the strengths of arbitrary TBRs
with no restriction for the orbital elements of the particle's orbit
but taking circular and coplanar orbits for the two planets.
Our results are roughly in agreement with previous theoretical and numerical studies of TBRs involving Jupiter-Saturn
for the planar case and predict some new results. For example, our algorithm allowed us to show that some TBRs have asymmetric equilibrium points, nevertheless, it cannot discriminate if they are
stable or unstable.
We found that the resonance's  strength is proportional to $e^{q}$
in agreement with theoretical studies but also that for zero eccentricity orbits it is proportional
 to $(\sin i)^{z}$ with $z=q$ for even order
TBRs and $z=2q$ for odd order TBRs.

The algorithm allows to identify the strongest TBRs near a given semimajor axis for a given
set of elements $(e,i,\omega)$ of the test particle and following this procedure we  obtained an atlas of the strongest TBRs in the Solar System.
The main asteroid belt is crossed by several strong TBRs involving Jupiter and Saturn that appear as peaks in an histogram of proper semimajor axes, but immerse in a region with the lowest density of TBRs in the Solar System. Both borders of the asteroid belt are characterized by an increase in the density of relatively strong TBRs
reinforcing the claims that TBRs have a relevant role in the stability  of that region.
We also found a series of strong TBRs of the type n+1U-2N with n integer in the far TNR, specially at $\sim 262$,  $\sim 416$ and  $\sim 545$ au.
As illustration we showed some objects in TBRs involving other planets than Jupiter and Saturn.
This paper it is not a systematic  survey of objects in TBRs, we are confident that if a survey of this kind is done
with the help of our method it will identify
several minor bodies in TBRs involving pairs of planets other than Jupiter-Saturn.

The complete atlas and codes for computing  $\rho(\sigma)$ and  $\Delta\rho$ are available under request to the
corresponding author.

\bigskip

\textbf{Acknowledgments.}
Two anonymous reviewers contributed to improve substantially the original manuscript.
Partial support from PEDECIBA is acknowledged.


\newpage

\begin{table}
  \centering
\begin{tabular}{  r   r    r      r  }
  \hline
  $a$ (au)      &          resonance                      & $q$ & $\Delta \rho$ \\
 \hline
   0.7946  &   $  8  -7   V    +1   J    $ &   2   &   0.00448	 \\
   1.0980  &   $  7  -6   E    -1   J    $ &   0   &   0.00450	 \\
   1.1020  &   $  8  -7   E    +1   J    $ &   2   &   0.00238	 \\
   1.9389  &   $  1  -6   J    +4   S    $ &   1   &   0.00131	 \\
   2.1376  &   $  1  -5   J    +3   S    $ &   1   &   0.00260	 \\
   2.3031  &   $  1  -5   J    +4   S    $ &   0   &   0.00121	 \\
   2.3967  &   $  1  -4   J    +2   S    $ &   1   &   0.00288	 \\
   2.4493  &   $  2  -9   J    +7   S    $ &   0   &   0.00127	 \\
   2.5026  &   $  2  -8   J    +5   S    $ &   1   &   0.00151	 \\
   2.6211  &   $  1  -4   J    +3   S    $ &   0   &   0.00282	 \\
   2.6845  &   $  2  -7   J    +4   S    $ &   1   &   0.00317	 \\
   2.7518  &   $  1  -3   J    +1   S    $ &   1   &   0.00161	 \\
   2.8266  &   $  2  -7   J    +5   S    $ &   0   &   0.00573	 \\
   2.8506  &   $  3  -9   J    +4   S    $ &   2   &   0.00107	 \\
   2.9034  &   $  2  -6   J    +3   S    $ &   1   &   0.00538	 \\
   2.9587  &   $  3  -9   J    +5   S    $ &   1   &   0.00248	 \\
   3.0155  &   $  3  -8   J    +3   S    $ &   2   &   0.00171	 \\
   3.0777  &   $  1  -3   J    +2   S    $ &   0   &   0.02678	 \\
   3.1406  &   $  3  -8   J    +4   S    $ &   1   &   0.00489	 \\
   3.1732  &   $  2  -5   J    +2   S    $ &   1   &   0.00288	 \\
   3.2794  &   $  3  -8   J    +5   S    $ &   0   &   0.00614	 \\
   3.3336  &   $  4  -9   J    +3   S    $ &   2   &   0.00241	 \\
   3.3933  &   $  2  -5   J    +3   S    $ &   0   &   0.01548	 \\
   3.4160  &   $  5 -11   J    +4   S    $ &   2   &   0.00251	 \\
   3.4532  &   $  4  -9   J    +4   S    $ &   1   &   0.00600	 \\
   3.5176  &   $  3  -7   J    +4   S    $ &   0   &   0.01068	 \\
   3.5842  &   $  4  -9   J    +5   S    $ &   0   &   0.00556	 \\
   3.6059  &   $  3  -6   J    +2   S    $ &   1   &   0.00562	 \\
   3.6822  &   $  5 -10   J    +4   S    $ &   1   &   0.00619	 \\
   3.7408  &   $  5  -9   J    +2   S    $ &   2   &   0.00490	 \\
   3.7513  &   $  6 -11   J    +3   S    $ &   2   &   0.00219	 \\
   3.8045  &   $  1  -2   J    +1   S    $ &   0   &   0.02768	 \\
   3.8846  &   $  4  -7   J    +2   S    $ &   1   &   0.01427	 \\
   3.9123  &   $  3  -5   J    +1   S    $ &   1   &   0.01196	 \\
   3.9727  &   $  4  -8   J    +5   S    $ &   1   &   0.00549	 \\
  \hline	 									
\end{tabular}	
 \caption{The strongest resonances with $\Delta\rho > 0.001$ in intervals of 0.05 au from 0 to 4 au
calculated assuming $e=0.15, i=6^{\circ}, \omega=60^{\circ}$.}\label{strongest10}					
\end{table}

\newpage

\begin{table}
  \centering
 \begin{tabular}{  r   r    r      r  }
  \hline
  $a$ (au)      &          resonance                      &  $q$  & $\Delta \rho$ \\
 \hline
    4.9242  &   $  7  -8   J  +1   S  $  &   0  &   0.29596     \\
    5.7240  &   $  3  -3   J  +1   S  $  &   1  &   3.37943     \\
    6.5919  &   $  2  -1   J  -1   S  $  &   0  &   0.20865     \\
    7.3022  &   $  3  -1   J  -2   S  $  &   0  &   0.16873     \\
    8.8460  &   $  4  -1   J  -2   S  $  &   1  &   0.30202     \\
    9.0482  &   $  6  +1   J  -9   S  $  &   2  &   0.63886     \\
   10.4640  &   $  4  -1   J  -1   S  $  &   2  &   0.45567     \\
   11.1722  &   $ 12  -1   J  -7   S  $  &   4  &   0.08574     \\
   12.1424  &   $  5  -1   J  -1   S  $  &   3  &   0.02895     \\	
   13.2836  &   $  9  -1   J  -3   S  $  &   5  &   0.00518     \\
   14.9629  &   $  1  +1   J  -3   S  $  &   1  &   0.00581     \\
   15.0977  &   $  3  +1   J  -4   S  $  &   0  &   0.00601     \\
   16.8947  &   $  4  -1   S  -2   U  $  &   1  &   0.00469     \\
   17.9233  &   $ 10  -1   J  -4   U  $  &   5  &   0.01298     \\
   18.0021  &   $  1  -1   J  +6   U  $  &   6  &   0.01544     \\
   19.0549  &   $  8  -1   J  -1   U  $  &   6  &   0.00881     \\
   20.6628  &   $  1  +1   J  -8   U  $  &   6  &   0.01141     \\
   21.0578  &   $  7  -1   J  +1   U  $  &   7  &   0.04680     \\
   22.1112  &   $ 10  -1   J  -1   U  $  &   8  &   0.00529     \\
   23.7521  &   $  2  +1   J  -3   S  $  &   0  &   0.00482     \\
   26.8747  &   $  5  -1   J  +8   N  $  &  12  &   0.00128     \\
   27.9823  &   $  8  -1   J  +5   N  $  &  12  &   0.00954     \\
   28.5246  &   $ 11  -1   J  +2   N  $  &  12  &   0.03031     \\
   29.4209  &   $  2  +1   J -16   N  $  &  13  &   0.00458     \\
   30.9517  &   $ 10  -1   S  -4   N  $  &   5  &   0.00198     \\
   31.8850  &   $ 13  -1   J  +2   N  $  &  14  &   0.00322     \\
   32.8328  &   $ 17  -1   J  -1   N  $  &  15  &   0.01268     \\
   33.2439  &   $ 15  -1   J  +1   N  $  &  15  &   0.05002     \\
   34.1081  &   $ 18  -1   J  -1   N  $  &  16  &   0.00354     \\	
  262.04  &   $  1  +1   U  -2   N  $  &   0  &   0.01373     \\
  415.96  &   $  2  +1   U  -2   N  $  &   1  &   0.00258     \\
  \hline	 									
\end{tabular} 
\caption{The strongest resonances with $\Delta\rho > 0.001$ in intervals of 1 au from 4 to 1000 au
calculated assuming $e=0.15, i=6^{\circ}, \omega=60^{\circ}$. }\label{strongest1000}							
\end{table}

\begin{table}
  \centering
 \begin{tabular}{  r   r    r      r  }
  \hline
  $a$ (au)      &          resonance                      & $q$  & $\Delta \rho$ \\
 \hline
   0.5838  &   $  5  -7   V  +2   J   $  &     0  &    0.00011	 \\
   0.6993  &   $  1  -1   V  -1   J   $  &     1  &    0.00036	 \\
   0.7946  &   $  8  -7   V  +1   J   $  &     2  &    0.00448	 \\
   0.8057  &   $  6  -5   V  -2   J   $  &     1  &    0.00054	 \\
   0.9817  &   $  3  -3   E  -1   J   $  &     1  &    0.00086	 \\
   1.0980  &   $  7  -6   E  -1   J   $  &     0  &    0.00450	 \\
   1.1020  &   $  8  -7   E  +1   J   $  &     2  &    0.00238	 \\
   1.2415  &   $  3  -2   E  -2   J   $  &     1  &    0.00039	 \\
   1.3381  &   $  8  -5   E  -2   J   $  &     1  &    0.00039	 \\
   1.4128  &   $  7  -4   E  -2   J   $  &     1  &    0.00039	 \\
   1.5040  &   $  2  -1   E  -1   J   $  &     0  &    0.00059	 \\
   1.6703  &   $  9  -8   M  +1   J   $  &     2  &    0.00037	 \\
   1.7452  &   $  5  -2   E  -2   J   $  &     1  &    0.00032	 \\
   1.8748  &   $  3  -1   E  -2   J   $  &     0  &    0.00056	 \\
   1.9708  &   $  3  -1   E  -1   J   $  &     1  &    0.00019	 \\
   2.1681  &   $  4  -1   E  -3   J   $  &     0  &    0.00024	 \\
   2.2712  &   $  4  -1   E  -2   J   $  &     1  &    0.00030	 \\
   2.4090  &   $  5  -1   E  -4   J   $  &     0  &    0.00010	 \\
   2.5159  &   $  5  -1   E  -3   J   $  &     1  &    0.00019	 \\
   2.6377  &   $  3  -1   M  -2   J   $  &     0  &    0.00011	 \\
   2.7204  &   $  6  -1   E  -4   J   $  &     1  &    0.00012	 \\
  \hline	 									
\end{tabular}	 									
 \caption{The strongest resonances with $\Delta\rho > 0.0001$  involving terrestrial and jovian planets in intervals of 0.1 au from 0 to 4 au
calculated assuming $e=0.15, i=6^{\circ}, \omega=60^{\circ}$. }\label{terrjov}
\end{table}

\begin{table}
  \centering
 \begin{tabular}{  r   r    r      r  }
  \hline
  $a$ (au)      &          resonance                      &  $q$  & $\Delta \rho$ \\
 \hline
     0.4249 &   $   3  -3  Me  +1   V  $ &   1 &   0.00002  \\
     0.5822 &   $   1  -2   V  +1   E  $ &   0 &   0.00004  \\
     0.6885 &   $   8  -8   V  -1   E  $ &   1 &   0.00016  \\
     0.7949 &   $   7  -3   V  -5   E  $ &   1 &   0.00042  \\
     0.8341 &   $   2  -1   V  -1   E  $ &   0 &   0.00019  \\
     0.9602 &   $   4  -2   V  -1   E  $ &   1 &   0.00014  \\
     1.0998 &   $   4  -4   E  +1   M  $ &   1 &   0.00017  \\
     1.1016 &   $  12  -7   V  +1   E  $ &   6 &   0.00070  \\
     1.2390 &   $   5  -1   V  -2   E  $ &   2 &   0.00004  \\
     1.3240 &   $   4  -1   V  -1   E  $ &   2 &   0.00002  \\
     1.6774 &   $  11  -4   E  -2   M  $ &   5 &   0.00003  \\
     1.9248 &   $   1  +1   V  -2   E  $ &   0 &   0.00007  \\
     3.0554 &   $   2  +1   V  -2   E  $ &   1 &   0.00001  \\
  \hline	 								
\end{tabular}	 								
 \caption{The strongest resonances with $\Delta\rho > 0.00001$  involving only terrestrial planets in intervals of 0.1 au
calculated assuming $e=0.15, i=6^{\circ}, \omega=60^{\circ}$.}\label{onlyterr}			
\end{table}

\clearpage

\end{document}